\documentclass[aps,pra,twocolumn,superscriptaddress,nofootinbib,floatfix]{revtex4-2}

\usepackage[colorlinks=true, linkcolor=blue, citecolor=blue, urlcolor=blue]{hyperref}

\usepackage{amsmath,amssymb,braket,bm}
\usepackage{hyperref,bbm,times,mathtools}
\usepackage{xcolor}
\usepackage{soul}

\newcommand{\Tr}{\mathrm{Tr}}

\begin{document}

\title{Energy-space quantum walks: Thermalization without state convergence}

\author{Alana Spak dos Santos}
\affiliation{Department of Physics, Federal University of Paran\'a, P.O. Box 19044, Curitiba 81531-980, Paran\'a, Brazil}

\author{Renato Moreira Angelo}
\affiliation{Department of Physics, Federal University of Paran\'a, P.O. Box 19044, Curitiba 81531-980, Paran\'a, Brazil}

\date{\today}

\begin{abstract}
We introduce energy-space quantum walks as a minimal framework to investigate equilibration, thermalization, and irreversibility from an effective-dynamics perspective. By mapping the configuration space of a walk onto a ladder of energy eigenlevels, we reinterpret thermalization as transport in energy space, independently of microscopic system--bath details. At the classical level, the resulting birth--death--lazy dynamics leads to equilibration of the energy distribution and, under suitable conditions, to a Gibbs stationary state. We then embed this dynamics into a unitary, collision-assisted model in which coherence is controlled by a single parameter. A central result is a structural decoupling between population dynamics and coherence generation: while the populations evolve according to the classical process and relax to the Gibbs distribution, the full quantum state exhibits a persistent coherence-induced deviation from the thermal manifold. This establishes a minimal scenario of thermalization without state convergence, where equilibration occurs at the level of populations but not at the level of the full density operator. We quantify this effect using the thermal distance to the Gibbs state and derive perturbative bounds that relate the long-time deviation to classical transport properties. Our results show that coherence acts as a controllable and quantitatively bounded source of nonthermal behavior, providing a clear separation between classical equilibration and genuinely quantum corrections.
\end{abstract}

\maketitle

\section{Introduction}

Thermalization and equilibration in quantum systems are routinely discussed in terms of microscopic unitary dynamics on enlarged Hilbert spaces, where effective irreversibility arises from coarse-graining and the neglect of environmental degrees of freedom \cite{Polkovnikov2011,Eisert2015,ShortFarrelly2012}. Within the theory of open quantum systems, such emergent irreversibility and entropy production are commonly described by Markovian quantum dynamical semigroups derived from system--bath models \cite{BreuerPetruccione2002,Davies1974,Spohn1978}. Despite this standard narrative, thermalization is often implicitly associated with convergence of the full quantum state to a Gibbs form.

However, many qualitative features of thermal behavior depend less on microscopic detail than on a few structural properties of the effective dynamics---particularly on how probability and coherence are redistributed across energy levels. This observation motivates the search for \emph{effective} dynamical descriptions in which energy itself plays the role of a synthetic dimension, allowing thermalization to be viewed as a transport phenomenon in \emph{energy space}. Closely related viewpoints have been developed in quantum thermodynamics and in resource-theoretic approaches to irreversibility, where structural and information-theoretic constraints take center stage \cite{Goold2016,Landi2021}.

From a complementary perspective, equilibration and thermalization in closed quantum systems have been rationalized in terms of typicality arguments and the eigenstate thermalization hypothesis (ETH), according to which individual energy eigenstates of generic many-body Hamiltonians already encode thermal properties for local observables \cite{Deutsch1991,Srednicki1994,Reimann2008,Popescu2006,Linden2009,Gogolin2016,Dalessio2016}. Within this paradigm, entanglement between subsystems plays a central role in the emergence of locally thermal behavior, while decoherence mechanisms stabilize effective classicality and suppress quantum recurrences \cite{Zurek2003,Schlosshauer2007}. More recently, quantum coherence itself has been identified as a resource with operational and thermodynamic significance, leading to a refined understanding of nonequilibrium processes and their limitations \cite{Baumgratz2014,Streltsov2017,Lostaglio2015}. Taken together, these insights reinforce the view that thermal behavior is governed by structural features of quantum dynamics---such as spectral properties, transport, and dephasing---rather than by specific microscopic reservoir models.

In parallel, quantum walks have emerged as a versatile framework to investigate coherent transport, interference, and the role of noise, with applications ranging from quantum information processing to energy transport in complex networks \cite{Aharonov1993,Kempe2003,Venegas2012}. More recently, quantum-walk-inspired and related network-based dynamics have been used to address thermodynamic questions, including the interplay between coherence and transport efficiency, environment-assisted quantum transport, and entropy production in noisy systems \cite{PlenioHuelga2008,Rebentrost2009,Whitfield2010,Romanelli2014}. These developments suggest that quantum walks provide a natural language to bridge coherent dynamics and effective thermodynamic behavior.

Here we take this perspective one step further by promoting the walker’s configuration space to a ladder of energy eigenlevels and interpreting the dynamics as transport in energy space. Specifically, we introduce \emph{energy-space quantum walks} as a minimal framework to investigate equilibration, the emergence of Gibbs stationary states, and irreversibility as properties of effective dynamics. The central idea is to reinterpret the ``position'' register of a walk as a ladder of energy eigenstates $\{\ket{n}\}_{n\in\mathbb{N}}$, with transitions $\ket{n}\!\leftrightarrow\!\ket{n\pm1}$ representing elementary energy exchanges. This viewpoint is naturally realized in modern settings such as synthetic dimensions and Floquet systems, where absorption and emission processes induce transport along effective energy axes \cite{Oka2005,RudnerLindner2020,Braiman2023}.

Our approach deliberately focuses on effective dynamics and on the identification of minimal structural ingredients required for thermal behavior. Rather than committing to a specific microscopic system--bath model, we ask which properties of an energy-space dynamics are sufficient to yield equilibration and, under suitable conditions, Gibbsian energy distributions. In this setting, localized breaks of translational invariance in energy space play a crucial role, as they allow coherence to be generated in a confined region while classical transport redistributes probability across the spectrum.

We show that classical birth--death--lazy processes on an energy ladder arise as the fully decohered limit of an underlying unitary, collision-assisted dynamics. Within this construction, a central structural result emerges: population dynamics and coherence generation become decoupled. Equivalently, the reduced dynamics acquires a block structure in the energy eigenbasis, in which the diagonal sector evolves autonomously according to a classical Markov process, while coherence is generated in the off-diagonal sector without back-action on the populations. As a consequence, equilibration at the level of populations does not imply convergence of the full quantum state whenever coherence is injected without back-action on the diagonal sector.

More generally, the perspective adopted in this work does not rely on a specific microscopic system--bath model. Rather, the behavior observed here arises in quantum dynamics with a block structure in the energy eigenbasis, where the diagonal sector evolves autonomously---typically as a classical Markov process---while coherence is generated in the off-diagonal sector without feedback onto the populations.

The energy-space quantum walk introduced in this work provides a minimal and physically transparent realization of this structure. By embedding a classical birth--death--lazy process into a unitary collision model, we explicitly construct a setting in which classical equilibration in energy space coexists with persistent quantum coherence. This allows us to identify and quantify a minimal scenario of \emph{thermalization without state convergence}, where irreversibility and Gibbsian behavior emerge at the level of populations, while the full density operator remains at a finite distance from the thermal manifold.

While the present construction is formulated in discrete time, its relation to continuous-time descriptions is nontrivial. In particular, the mechanism identified here relies on coherence injection localized in energy space and therefore does not, in general, straightforwardly map onto standard Lindblad or Davies generators. Whether an appropriate continuous-time limit can be defined remains an open question.

The paper is organized as follows: in Sec.~\ref{sec:classical_model} we introduce the classical energy-space dynamics and analyze its stationary states; in Sec.~\ref{sec:temporal_equilibrium} we discuss temporal equilibration and diagnostics of thermalization; in Sec.~\ref{sec:coin_qw_energy} we present the unitary, collision-assisted embedding and analyze the role of coherence; in Sec.~\ref{sec:numerics} we illustrate the dynamics numerically; and in Sec.~\ref{sec:conclusions} we conclude.

\section{Minimal stochastic dynamics in energy space}
\label{sec:classical_model}

We begin by introducing a minimal effective description of transport in energy space, corresponding to the fully incoherent (classical) limit of the energy-space quantum walk developed in subsequent sections. Although the dynamics is purely stochastic at the level of populations, it is convenient to formulate it from the outset in the language of linear maps acting on operators. This representation anticipates the quantum generalization with a coin degree of freedom and provides a unified framework to discuss fixed points, unitality, and irreversibility.

\subsection{Energy-space dynamics as a linear map}

We consider a system with a nondegenerate ladder of energy eigenstates $\{\ket{n}\}_{n\ge 0}$ and associated projectors $M_n=\ket{n}\!\bra{n}$. 
States that are diagonal in the energy basis form a convex set
\begin{equation}
\!\!\mathcal D=\left\{\rho=\sum_{n\ge 0} \wp(n)\,M_n : \wp(n)\ge 0,\;\sum_{n\ge 0}\wp(n)=1\right\}\!,
\end{equation}
which is in one-to-one correspondence with the set of classical probability distributions over the nonnegative integers.

We define a linear, trace-preserving map $\Phi:\mathcal D\to\mathcal D$ by its action on the basis projectors,
\begin{equation}
\Phi(M_n)=
\begin{cases}
(p_-+p_0) M_0 + p_+ M_1, & n=0,\\[4pt]
p_- M_{n-1} + p_0 M_n + p_+ M_{n+1}, & n\ge 1,
\end{cases}
\label{eq:Phi_action}
\end{equation}
where $p_+,p_-,p_0\ge0$ satisfy $p_++p_-+p_0=1$. These parameters represent the probabilities for upward transitions ($p_+$), downward transitions ($p_-$), and ``lazy'' steps ($p_0$), corresponding respectively to absorption, emission, and elastic processes (no change in energy).

By linearity, the evolution of an arbitrary diagonal state is given by
\begin{equation}
\rho_{t+1}=\Phi(\rho_t)=\sum_{n\ge 0} \wp_t(n)\,\Phi(M_n).
\end{equation}
The induced dynamics for the populations $\wp_t(n)=\Tr[M_n\rho_t]$ takes the form of a birth--death--lazy process with an additional lazy channel,
\begin{subequations}
\begin{align}
\wp_{t+1}(0) &= (p_-+p_0)\wp_t(0) + p_-\,\wp_t(1), \label{eq:recurrence_0}\\
\wp_{t+1}(n) &= p_-\,\wp_t(n+1) + p_0\,\wp_t(n) + p_+\,\wp_t(n-1),
\label{eq:recurrence_n}
\end{align}
\end{subequations}
for $n \ge 1$. The boundary condition at $n=0$ enforces the existence of a ground state and prevents transitions to negative energies.

\subsection{Operator-sum representation and nonunitality}

The map $\Phi$ admits a Kraus representation that preserves diagonality in the energy basis. A convenient choice is
\begin{subequations}
\begin{align}
K_-^{\mathfrak{b}} & = \sqrt{p_-} \ket{0}\bra{0}, \\
K_- &= \sqrt{p_-} \sum_{n\ge 1} \ket{n-1}\bra{n}, \\
K_0 &= \sqrt{p_0}\,\mathbbm{1}, \\
K_+ &= \sqrt{p_+}\sum_{n\ge 0} \ket{n+1}\bra{n},
\end{align}
\end{subequations}
where the superscript $\mathfrak{b}$ denotes the boundary contribution at $n=0$. The map then reads
\begin{equation}
\Phi(\rho_t)= K_- \rho_t K_-^\dagger + K_0 \rho_t K_0^\dagger + K_+ \rho_t K_+^\dagger +  K_-^{\mathfrak{b}} \rho_t K_-^{\mathfrak{b}\,\dagger},
\end{equation}
with
\begin{equation}
K_-^\dagger K_- + K_0^\dagger K_0 + K_+^\dagger K_+ + K_-^{\mathfrak{b}\, \dagger} K_-^{\mathfrak{b}}= \mathbbm{1}.
\end{equation}

This representation shows that the stochastic energy-space dynamics can be viewed as a quantum channel that preserves diagonality, i.e., a classical Markov process embedded in a completely positive trace-preserving map (CPTP) map. 

The map $\Phi$ is nonunital whenever $p_+\neq p_-$, since
\begin{equation}
\Phi(\mathbbm{1}) = \mathbbm{1} + (p_- - p_+) M_0 \neq \mathbbm{1}.
\end{equation}
As a consequence, the maximally mixed state is not a fixed point of the dynamics. This nonunitality reflects the presence of a bias in energy space and leads to a nontrivial stationary state. In particular, the condition $p_+\neq p_-$ selects a preferred direction of transport, with mean increment $\mathbb{E}[\Delta n]=p_+-p_-$ per step. This bias induces a systematic flow of probability toward lower or higher energy levels and provides a simple structural signature of irreversibility at the level of the effective dynamics.

From an information-theoretic viewpoint, irreversibility can be characterized through the convergence of the state toward a stationary distribution. In the present setting, this motivates the use of distances and mixing times as diagnostics of equilibration and irreversibility, which will be introduced below.

\subsection{Stationary state and Gibbs form}

A stationary state $\rho_\infty\in\mathcal D$ satisfies $\Phi(\rho_\infty)=\rho_\infty$. At the level of populations, Eqs.~\eqref{eq:recurrence_0}--\eqref{eq:recurrence_n} imply
\begin{align}
\wp_\infty(0) &= (p_-+p_0)\wp_\infty(0)+p_-\,\wp_\infty(1),\nonumber\\
\wp_\infty(n) &= p_-\,\wp_\infty(n+1)+p_0\,\wp_\infty(n)+p_+\,\wp_\infty(n-1),\nonumber
\end{align}
for $n\ge1$. These stationary equations fully determine the population dynamics. It is convenient, though, to rewrite them in a form that highlights their transport structure. To this end, we introduce the probability current
\begin{equation}
J_n := p_+\,\wp_\infty(n)-p_-\,\wp_\infty(n+1),
\end{equation}
in terms of which the stationary condition becomes a discrete continuity equation,
\begin{equation}
J_{n-1}=J_n.
\end{equation}
Thus, the current is constant along the energy ladder. The boundary condition fixes $J_0=0$, implying $J_n=0$ for all $n$, which yields the local balance condition
\begin{equation}
p_-\,\wp_\infty(n+1)=p_+\,\wp_\infty(n).
\end{equation}

For $p_->p_+$, the local balance condition obtained from the stationary equations admits a unique normalizable solution (see Appendix~\ref{app:stationary_derivation}),
\begin{subequations}
\begin{align}
\wp_\infty(n) &= \left(\frac{p_+}{p_-}\right)^n \wp_\infty(0),\\
\wp_\infty(0) &= \frac{p_- - p_+}{p_-}.
\end{align}
\end{subequations}
This yields a geometric stationary distribution. For equally spaced energies, $E_n=nE$, this distribution can be written in Boltzmann form, $\wp_\infty(n)\propto e^{-\beta E_n}$, which leads to the identification $\frac{p_+}{p_-}=e^{-\beta(E_{n+1}-E_n)}$. In this case, the stationary state is Gibbsian with
\begin{equation}
T=\frac{E}{\ln(p_-/p_+)},
\end{equation}
showing that thermal equilibrium arises as a property of biased transport in energy space.

For equally spaced spectra, the stationary distribution reduces to the canonical Gibbs distribution for a linear energy ladder, $\wp_\beta(n)=(1-e^{-\beta E})e^{-\beta En}$, upon identifying $e^{-\beta E}=p_+/p_-$. This corresponds to the thermal occupation of a quantum harmonic oscillator. Thus, the effective dynamics reproduces Gibbs statistics without invoking an explicit thermal bath.

While the stationary distribution characterizes equilibrium at the level of populations, it does not generally describe the long-time behavior of the full quantum state. In this case, a complementary notion of equilibrium is provided by temporal averaging.

\section{Temporal equilibrium and equilibration times}
\label{sec:temporal_equilibrium}

In addition to stationary states defined as fixed points of the map $\Phi$, a natural and operational notion of equilibrium is provided by temporal averaging. Given an initial state $\rho_0\in\mathcal D$ and the discrete-time evolution $\rho_t=\Phi^t(\rho_0)$, we define the time-averaged state
\begin{equation}
\bar\rho_\tau \coloneqq \frac{1}{\tau}\sum_{t=0}^{\tau-1}\rho_t,
\qquad \bar\rho := \lim_{\tau\to\infty}\bar\rho_\tau,
\label{eq:cesaro}
\end{equation}
whenever the limit exists. The state $\bar\rho$ characterizes the long-time behavior of expectation values $\Tr(O\bar\rho)$ for any bounded observable $O$~\cite{Gogolin2016}.

For the present birth--death--lazy dynamics with $p_->p_+$, the map $\Phi$ admits a unique stationary state $\rho_\infty$. In this case, convergence $\rho_t\to\rho_\infty$ implies that the temporal equilibrium coincides with the stationary state,
\begin{equation}
\bar\rho=\rho_\infty .
\end{equation}
While this equivalence holds in the present incoherent regime, it will no longer hold automatically once coherent unitary walk dynamics are introduced. Temporal averaging then provides a robust notion of equilibrium in the presence of persistent oscillations or quasiperiodic dynamics.

We now quantify the approach to equilibrium by introducing suitable distance measures and equilibration times.

\subsection{Distances to equilibrium}
\label{sec:distances}

To quantify equilibration, we consider the trace distance between the evolving state and a reference equilibrium state. Since all states in the present section are diagonal in the energy basis, the trace distance
\begin{equation}
    d_1(\rho,\sigma)=\tfrac12\|\rho-\sigma\|_1,
    \label{eq:d1}
\end{equation}
coincides with the total-variation distance between the corresponding probability distributions. Given a target state $\rho_\star$ (either the stationary state $\rho_\infty$ or the temporal equilibrium $\bar\rho$), we analyze the behavior of $d_1(\rho_t,\rho_\star)$ as a function of time.

For numerical simulations one typically works with finite truncations of the energy ladder. In this setting, the approach to the stationary state is governed by the spectral gap of the associated transition matrix\footnote{For finite Markov chains, the spectral gap---defined as the difference between the largest and second-largest eigenvalues of the transition matrix---controls the asymptotic rate of convergence to the stationary distribution; see, e.g., \cite{Levin2009,Norris1997}.}, leading to an exponential convergence at long times. As the bias $p_- - p_+$ decreases, the gap narrows and equilibration becomes progressively slower. The relaxation timescale may also depend on properties of the initial distribution, such as its mean energy and spread in energy space.

\subsection{Approach to Gibbs states}

When the stationary state $\rho_\infty$ coincides with a Gibbs state $\rho_\beta$ (as in the equally spaced spectrum discussed above), equilibration to $\rho_\infty$ also constitutes thermalization in the usual sense. More generally, even when $\rho_\infty$ does not have an exact Gibbs form, one may quantify the proximity of $\rho_t$ to the Gibbs manifold by comparing it with the Gibbs state having the same instantaneous mean energy. We define $\rho_{\beta_t}$ as the Gibbs state satisfying
\begin{equation}\label{eq:<H>}
\Tr(H\rho_{\beta_t})=\Tr(H\rho_t),
\end{equation}
and introduce the thermal distance
\begin{equation}
d_{\mathrm{th}}(t):=d_1(\rho_t,\rho_{\beta_t}).
\label{eq:dth(t)}
\end{equation}
Although we quantify deviations from thermal behavior using the trace distance to the Gibbs state with matching mean energy, the qualitative features discussed here do not depend on this specific choice. In particular, the persistence of a nonvanishing asymptotic deviation is a structural consequence of the dynamics and is therefore expected to hold for other contractive measures of distance, such as the relative entropy.

For equally spaced spectra $E_n=nE$, the parameter $\beta_t$ can be obtained explicitly from the instantaneous mean energy. Let $\langle n\rangle_t := \sum_{n\ge0} n\,\wp_t(n)$ denote the instantaneous mean occupation number. The corresponding Gibbs distribution takes the geometric form
\begin{equation}
\wp_{\beta_t}(n)=(1-q_t)q_t^n,
\qquad
q_t=e^{-\beta_t E},
\end{equation}
which has mean occupation number
\begin{equation}
\langle n\rangle_{\beta_t}=\frac{q_t}{1-q_t}.
\end{equation}
Inverting this relation yields
\begin{equation}
q_t=\frac{\langle n\rangle_t}{1+\langle n\rangle_t},
\qquad
\beta_t=\frac{1}{E}\ln\!\left(1+\frac{1}{\langle n\rangle_t}\right).
\end{equation}

This diagnostic allows one to distinguish convergence to the stationary state of the effective dynamics from genuine thermal behavior. In particular, a state may approach $\rho_\infty$ while remaining at a finite distance from the Gibbs manifold. The quantities $d_1(\rho_t,\rho_\infty)$ and $d_{\mathrm{th}}(t)$ therefore provide complementary diagnostics of relaxation.

For states diagonal in the energy basis, the trace distance reduces to the total-variation distance between the corresponding populations,
\begin{equation}\label{eq:d_th}
d_{\mathrm{th}}(t)
= \tfrac12 \sum_{n\ge 0} \big|\wp_t(n)-\wp_{\beta_t}(n)\big|.
\end{equation}
Thus, $\beta_t$, $q_t$, and $\wp_{\beta_t}(n)$ are directly constructed from $\langle n\rangle_t$, enabling a straightforward evaluation of $d_{\mathrm{th}}(t)$ in numerical simulations.

\subsection{Level-dependent rates and local detailed balance}
\label{subsec:level_dependent}

The constant-rate dynamics introduced above provides the simplest setting in which a Gibbs fixed point emerges for an equally spaced spectrum. A natural generalization is to allow the transition probabilities to depend on the energy level, thereby capturing the fact that microscopic transition rates typically depend on energy gaps and density-of-states factors.

We consider maps $\Phi:\mathcal D\to\mathcal D$ defined on the projectors $M_n=\ket{n}\!\bra{n}$ by
\begin{equation}
\Phi(M_n)=p_-(n)\,M_{n-1}+p_0(n)\,M_n+p_+(n)\,M_{n+1},
\end{equation}
for $n\ge 1$, together with the boundary action
\begin{equation}
\Phi(M_0)=\big[p_-(0)+p_0(0)\big]M_0+p_+(0)\,M_{1},
\end{equation}
where $p_\pm(n),p_0(n)\in[0,1]$ satisfy $p_-(n)+p_0(n)+p_+(n)=1$. The term $p_0(n)$ represents a lazy step that ensures aperiodicity without affecting the stationary structure.

For diagonal states, the induced dynamics defines a birth--deat--lazy process with level-dependent rates. When the stationary current
\begin{equation}
J_n := \wp_\infty(n)\,p_+(n)-\wp_\infty(n+1)\,p_-(n+1)
\end{equation}
vanishes, the stationary distribution satisfies
\begin{equation}\label{eq:ratio}
\frac{\wp_\infty(n+1)}{\wp_\infty(n)}=\frac{p_+(n)}{p_-(n+1)}.
\end{equation}
Iterating this relation yields the product form
\begin{equation}
\wp_\infty(n)=\wp_\infty(0)\,
\prod_{k=0}^{n-1}\frac{p_+(k)}{p_-(k+1)},
\end{equation}
provided normalization holds.

Let $H=\sum_{n\ge0}E_n M_n$. A Gibbs state has populations $\wp_\beta(n)\propto e^{-\beta E_n}$. Comparing the ratio in Eq.~\eqref{eq:ratio} with $\wp_\beta(n+1)/\wp_\beta(n)=e^{-\beta(E_{n+1}-E_n)}$ shows that a sufficient condition for $\rho_\beta$ to be stationary is the local detailed-balance relation
\begin{equation}
\frac{p_+(n)}{p_-(n+1)}=e^{-\beta\big(E_{n+1}-E_n\big)}.
\label{eq:local_detailed_balance}
\end{equation}
Under this condition, the stationary state takes the Gibbs form for arbitrary spectra $\{E_n\}$.

This shows that the emergence of Gibbs stationary states is not tied to equally spaced spectra or constant rates, but rather to a structural constraint on the transition probabilities. The relation \eqref{eq:local_detailed_balance} is the discrete analogue of detailed-balance conditions arising in microscopic derivations of open-system dynamics, such as Davies generators, where transition rates satisfy $W_{n\to n+1}/W_{n+1\to n}=e^{-\beta(E_{n+1}-E_n)}$~\cite{BreuerPetruccione2002,Davies1974}.

The lazy probabilities $p_0(n)$ do not affect the stationary ratios: under the zero-current condition $J_n=0$ one obtains Eq.~\eqref{eq:ratio}, which is independent of $p_0(n)$. They do, however, influence relaxation times through the spectrum of the transition matrix (e.g., adding a lazy component shifts eigenvalues and reduces the spectral gap). Moreover, if $p_+(n)/p_-(n+1)\neq e^{-\beta(E_{n+1}-E_n)}$ for some $n$, the product-form stationary state cannot be written as $\wp_\infty(n)\propto e^{-\beta E_n}$, and thus deviates from a Gibbs distribution.

\subsection{Classical equilibration: numerical illustration}
\label{subsec:classical_equilibration}

To illustrate the equilibration mechanisms discussed above, we present numerical simulations of the incoherent energy-space dynamics on finite truncations of the ladder, $n=0,1,\dots,N$. The cutoff $N$ is taken sufficiently large that boundary effects at the upper edge are negligible on the timescales considered. Starting from an initial diagonal state $\rho_0$, the map $\Phi$ is iterated to generate the sequence $\rho_t=\Phi^t(\rho_0)$, with $t\in\mathbbm{N}$.

\begin{figure}
\centering
\includegraphics[scale=0.5]{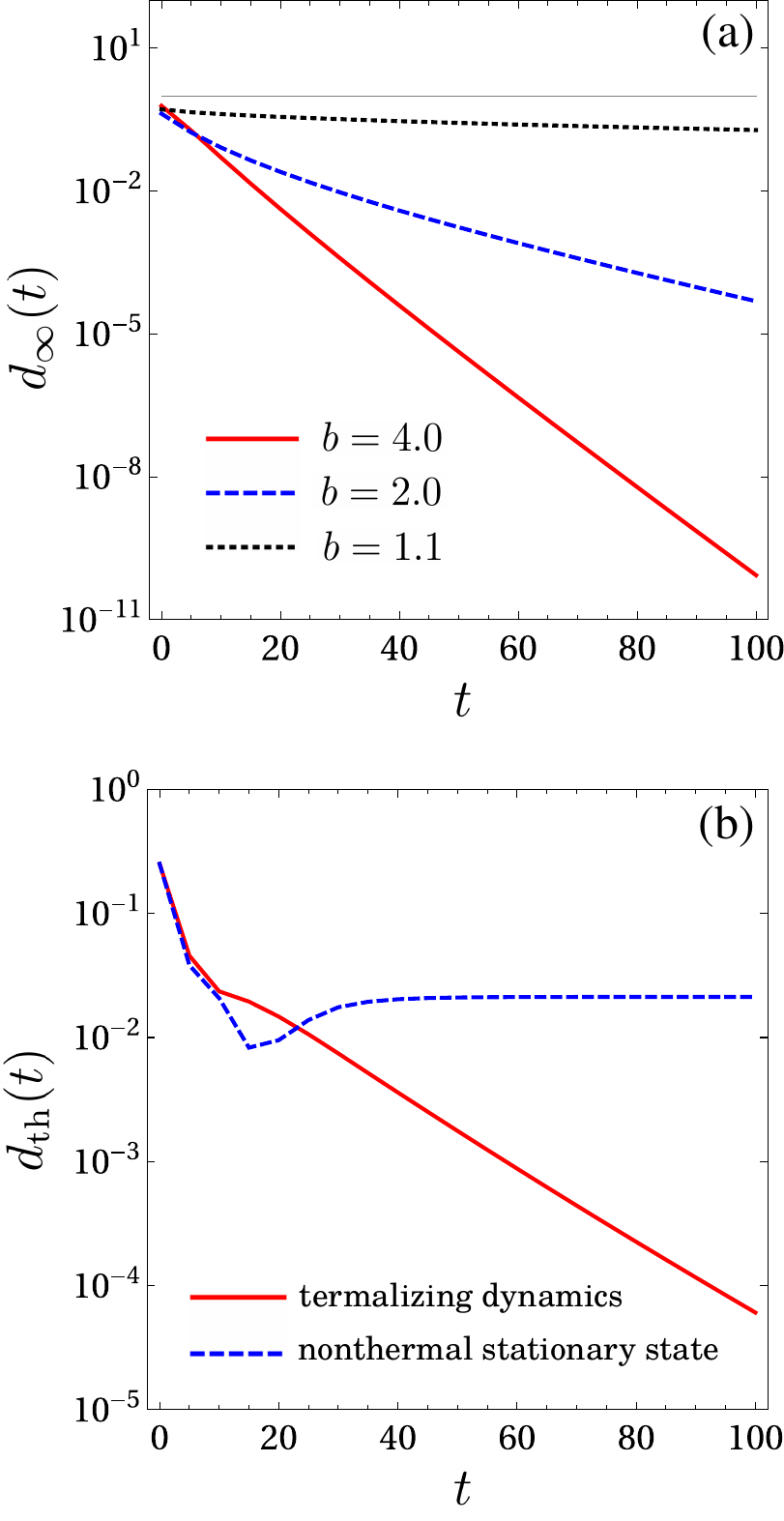}
\caption{Equilibration and thermalization in the incoherent energy-space dynamics for a system with equally spaced energies $E_n=nE$. The simulations use a truncated ladder $n=0,1,\dots,N$ with $N=50$, fixed lazy probability $p_0=0.1$, and a normalized discrete Gaussian initial distribution (``cold state'') $\wp_0(n)\propto e^{-(n-n_0)^2/(2\sigma^2)}$ with $n_0=2$ and $\sigma=2$.
(a) Time evolution of the distance to the stationary state, $d_\infty(t)=d_1(\rho_t,\rho_\infty)$, where $d_1$ is the trace distance defined in Eq.~\eqref{eq:d1}, for representative values of the bias parameter $b=p_-/p_+$ (solid red: $b=4$; dashed blue: $b=2$; dotted black: $b=1.1$). The decay slows down as $b\to1$, reflecting the reduced drift in energy space and the corresponding increase in relaxation times. (b) Distance to the Gibbs manifold, $d_{\mathrm{th}}(t)=d_1(\rho_t,\rho_{\beta_t})$, where $\rho_{\beta_t}$ is the Gibbs state with the same instantaneous mean energy as $\rho_t$. Constant transition rates (solid red curve) satisfy local detailed balance and lead to $d_{\mathrm{th}}(t)\to0$, indicating genuine thermalization. Level-dependent rates $p_+(n)=0.15+0.05/(1+n)$ and $p_-(n)=0.55-0.10/(2+n)$ (dashed blue), with $p_0(n)=1-p_+(n)-p_-(n)$, violate local detailed balance; the dynamics still equilibrates but approaches a nonthermal stationary state, so $d_{\mathrm{th}}(t)$ saturates at a finite value.}
\label{fig:1}
\end{figure}

We consider normalized discrete Gaussian initial distributions of the form
\begin{equation}
\wp_0(n)\propto e^{-(n-n_0)^2/(2\sigma^2)} .
\end{equation}
This family provides a flexible and physically transparent choice, allowing one to probe different initial energy regimes through the parameters $n_0$ and $\sigma$. In particular, the simulations shown in Fig.~\ref{fig:1} use a ``cold Gaussian state'', centered close to the ground state ($n_0=2$) with width $\sigma=2$, from which relaxation towards the stationary distribution can be clearly visualized.

We monitor two distances derived from the trace distance $d_1(\rho,\sigma)$ defined in Eq.~\eqref{eq:d1}. The first measures convergence to the stationary state, $d_\infty(t):=d_1(\rho_t,\rho_\infty)$, while the second quantifies the distance to the Gibbs manifold, $d_{\mathrm{th}}(t):=d_1(\rho_t,\rho_{\beta_t})$, where $\rho_{\beta_t}$ denotes the Gibbs state whose mean energy matches that of $\rho_t$.

Figure~\ref{fig:1}(a) shows the time evolution of $d_\infty(t)$ for representative values of the bias $b\equiv p_-/p_+$. In all cases the long-time relaxation is well described by an exponential decay, consistent with the spectral-gap estimate. As the bias approaches the unbiased regime $b\to1$, the drift in energy space decreases and equilibration becomes progressively slower. Although the simulations shown correspond to a cold Gaussian initial state, the qualitative behavior is robust with respect to the choice of initial distribution. In particular, if one starts from a ``hot Gaussian state'' centered at higher energies, one typically observes an initial transient regime before entering the same exponential decay controlled by the spectral gap. This transient reflects the larger distance that the distribution must traverse in energy space before reaching the stationary region of the ladder. Apart from this short-time effect, the long-time relaxation dynamics is essentially independent of the initial state.

Panel (b) of Fig.~\ref{fig:1} illustrates the behavior of the thermal distance $d_{\mathrm{th}}(t)$. At each time step we construct the Gibbs state $\rho_{\beta_t}$ whose mean energy matches that of the evolving state $\rho_t$, i.e., $\Tr(H\rho_{\beta_t})=\Tr(H\rho_t)$, as described in Sec.~\ref{sec:temporal_equilibrium}. For equally spaced energies $E_n=nE$, this condition uniquely determines $\beta_t$ through the instantaneous mean occupation number $\langle n\rangle_t$ [see Eqs.~\eqref{eq:<H>}-\eqref{eq:d_th}]. The quantity $d_{\mathrm{th}}(t)$ therefore measures the distance between the instantaneous population distribution and the corresponding Gibbs distribution.

When the transition rates are constant and the spectrum is equally spaced, the dynamics satisfies local detailed balance and the stationary state is Gibbsian, so that $d_{\mathrm{th}}(t)\to0$. By contrast, when the transition probabilities depend on the energy level and violate the local detailed-balance condition, the system still equilibrates to a stationary state but does not thermalize: $d_{\mathrm{th}}(t)$ saturates to a nonzero value. This provides a direct operational signature that the stationary state does not belong to the Gibbs manifold, in agreement with the violation of the local detailed-balance condition discussed in Sec.~\ref{subsec:level_dependent}. This comparison highlights the distinction between equilibration to the stationary state of the effective dynamics and genuine thermalization in the Gibbs sense.

\section{Collision-assisted quantum walks in energy space}
\label{sec:coin_qw_energy}

Discrete-time quantum walks provide a natural framework to describe coherent transport processes on graphs and lattices \cite{Aharonov1993,Ambainis2001,Kempe2003,Venegas2012}. In the coined formulation, an internal degree of freedom controls conditional transitions of the walker along the underlying structure, leading to interference-driven dynamics that can differ qualitatively from classical random walks.

Here we adopt a different and more physically grounded route, based on repeated interactions (collision models), which provide a minimal microscopic description of open-system dynamics and quantum thermodynamics~\cite{AttalPautrat2006,Strasberg2017,Ciccarello2017,Ciccarello2022PR}. In this approach, the system interacts sequentially with a stream of ancillary degrees of freedom, each prepared in a fixed state. This construction yields a unitary embedding of effective stochastic dynamics while maintaining a transparent physical interpretation.

We now extend the incoherent energy-space dynamics introduced in Sec.~\ref{sec:classical_model} by coupling the system to a sequence of three-level ancillary systems that encode the elementary transition channels. This allows us to investigate how quantum coherence in the ancilla affects transport, equilibration, and thermalization in energy space.

\subsection{Hilbert space and unitary step}
\label{subsec:coin_unitary_step}

To embed the classical energy-space dynamics into a unitary framework, we introduce an auxiliary three-level system (ancilla) that controls the elementary transitions along the energy ladder. Because the stochastic model of Sec.~\ref{sec:classical_model} admits three possible outcomes---upward, downward, or lazy transitions---we consider a three-dimensional ancillary space, which naturally reproduces the full birth--death--lazy process in an appropriate incoherent limit.

The total Hilbert space for a single interaction is therefore $\mathcal{H}=\mathcal{H_\mathcal{E}}\otimes\mathcal{H_\mathcal{A}}$, with $\mathcal{H_\mathcal{E}}=\mathrm{span}\{\ket{n}\}_{n\ge0}$ and $\mathcal{H_\mathcal{A}}=\mathrm{span}\{\ket{+},\ket{0},\ket{-}\}$. The ancillary states $\ket{+}$, $\ket{0}$, and $\ket{-}$ correspond respectively to upward, lazy, and downward transitions in energy space. In this sense, the ancilla plays a role analogous to the coin degree of freedom in standard quantum walks, encoding the directional channels that govern motion along the energy ladder.

The joint system--ancilla evolution is implemented by a unitary operator $U_\mathcal{S}$ that transports amplitudes along the energy ladder conditionally on the ancilla state. To write it compactly we introduce the ladder operators
\begin{subequations}
\begin{align}
S_+ &\coloneqq \sum_{n\ge0}\ket{n+1}\!\bra{n}, \\
S_- &\coloneqq S_+^\dagger=\sum_{n\ge1}\ket{n-1}\!\bra{n}.
\end{align}
\end{subequations}
These satisfy $S_-S_+=\mathbbm{1}_\mathcal{E}$ and $S_+S_-=\mathbbm{1}_\mathcal{E}-\ket{0}\!\bra{0}$, so that the deviation from unitarity of the unilateral shift is entirely encoded in the boundary projector $P_0\coloneqq\ket{0}\!\bra{0}$. Let $Q_0\coloneqq \mathbbm{1}_\mathcal{E}-P_0$ denote its orthogonal complement. The conditional shift can then be written as
\begin{align}
U_\mathcal{S}&=
S_+\otimes\ket{-}\!\bra{+}+S_-\otimes\ket{+}\!\bra{-} \nonumber \\
&+ Q_0\otimes\ket{0}\!\bra{0} +
P_0\otimes V_{\mathfrak{b}},
\label{eq:U_shift}
\end{align}
where $V_{\mathfrak{b}}$ is a unitary acting on the boundary ancillary subspace $\mathrm{span}\{\ket{0},\ket{-}\}\subset \mathcal {H_A}$ that closes the dynamics at $n=0$. For the simple choice
\begin{align}
V_{\mathfrak{b}}=\ket{0}\!\bra{0}+\ket{-}\!\bra{-},
\end{align}
the shift operator reduces to
\begin{align}
U_\mathcal{S}
&=S_+\otimes\ket{-}\!\bra{+}+S_-\otimes\ket{+}\!\bra{-} \nonumber \\
& + \mathbbm{1}_\mathcal{E}\otimes\ket{0}\!\bra{0}
+ P_0\otimes \ket{-}\!\bra{-}.
\end{align}
The shift operator has the structure of a flip-flop quantum walk, a standard construction in the quantum-walk literature \cite{Portugal2013}. In this formulation the ancillary states encode directional transport channels, while the shift operator propagates these channels along the underlying graph. This structure naturally accommodates boundary conditions on semi-infinite domains, such as the reflecting boundary adopted here. In the present context, these channels acquire a direct physical meaning: they correspond to elementary upward, downward, or lazy transitions along the energy ladder, so that the dynamics describes transport in energy space.

Equation~\eqref{eq:U_shift} shows that the bulk dynamics implements a scattering-type transport of the directional channels $\ket{\pm}$ along the ladder, while the lazy channel remains localized. Explicitly, for $n\ge1$, $\ket{n,+}\mapsto\ket{n+1,-}$, $\ket{n,0}\mapsto\ket{n,0}$, and $\ket{n,-}\mapsto\ket{n-1,+}$, whereas the boundary action at $n=0$ is determined by $V_{\mathfrak{b}}$, which acts unitarily on the subspace $\mathrm{span}\{\ket{0},\ket{-}\}$. This construction ensures that $U_\mathcal{S}$ is unitary while implementing transport on the semi-infinite energy ladder.

In the collision-model picture, at each time step the system interacts with a freshly prepared ancilla. The state of the ancilla encodes the transition statistics. We consider a one-parameter family of ancillary states
\begin{equation}
\rho_{\mu}=(1-\mu)\rho_{\mathrm{inc}}+\mu\,\ket{\chi}\!\bra{\chi},
\qquad 0\le\mu\le1,
\end{equation}
where
\begin{subequations}
\begin{align}
\rho_{\mathrm{inc}}
&=p_+\ket{+}\!\bra{+}
+p_0\ket{0}\!\bra{0}+p_-\ket{-}\!\bra{-},\\
\ket{\chi}
&=\sqrt{p_+}\ket{+}+\sqrt{p_0}\ket{0}+\sqrt{p_-}\ket{-}.
\end{align}
\end{subequations}
with $p_++p_0+p_-=1$. By construction, all states $\rho_\mu$ share the same diagonal $(p_+,p_0,p_-)$, so that $\mu$ controls only the amount of coherence between the transition channels. In line with the classical analysis of Sec.~\ref{sec:classical_model}, we focus on initial states that are diagonal in the energy basis, so that all coherence observed in the dynamics is generated by the evolution itself. This choice allows us to isolate the role of boundary-induced coherence in a controlled manner. The extension to initially coherent states is straightforward in principle, but leads to a less transparent separation between classical and quantum contributions.

\subsection{Reduced dynamics in energy space}
\label{subsec:reduced_dynamics}

The effective dynamics of the energy degree of freedom is obtained by tracing out the ancilla after each interaction. For an initial state $\rho_0$ of the system, a single step of the evolution is described by the CPTP map
\begin{equation}
\Phi_\mu(\rho_0)=
\operatorname{Tr}_\mathcal{A}\!\left[
U_\mathcal{S}\,(\rho_0\otimes\rho_{\mu})\,U_\mathcal{S}^\dagger
\right],
\label{eq:reduced_map}
\end{equation}
where $\rho_\mu$ denotes the ancilla state. The parameter $\mu$ controls the amount of coherence in the ancilla, and hence the degree to which the resulting dynamics departs from a classical stochastic evolution.

To make this connection explicit, we first consider the incoherent case $\rho_{\mathrm{inc}}=\rho_{\mu=0}$ and use the decomposition of $U_\mathcal{S}$ in Eq.~\eqref{eq:U_shift}. In the absence of coherence between the channels, no interference terms arise, and a direct calculation yields
\begin{equation}\label{eq:Phi_0}
\Phi_0(\rho_0)=p_+\,S_+\rho_0 S_+^\dagger+p_0\,\rho_0+p_- \left(S_-\rho_0 S_-^\dagger +P_0\rho_0 P_0\right),
\end{equation}
which coincides with the birth--death--lazy dynamics introduced in Sec.~\ref{sec:classical_model} for states $\rho_0$ that are diagonal in the energy basis.

For $\mu>0$, coherence between the states $\ket{+}$, $\ket{0}$, and $\ket{-}$ generates interference terms in Eq.~\eqref{eq:reduced_map}, coupling different transition channels and leading to genuinely quantum corrections. The map $\Phi_\mu$ therefore provides a continuous interpolation between the classical stochastic evolution at $\mu=0$ and a coherence-driven dynamics for $\mu>0$.

An explicit expression for $\Phi_\mu$ can be obtained by expanding Eq.~\eqref{eq:reduced_map} in the ancilla basis $\{\ket{+},\ket{0},\ket{-}\}$. Writing
$\rho_\mu=\sum_{c,c'\in\{+,0,-\}} r_{cc'}\,\ket{c}\!\bra{c'}$, with $r_{cc}=p_c$ and $\sum_c p_c=1$, and recalling that
$\rho_\mu=(1-\mu)\rho_{\mathrm{inc}}+\mu\ket{\chi}\!\bra{\chi}$,
with $\rho_{\mathrm{inc}}=\sum_c p_c\ket{c}\!\bra{c}$ and
$\ket{\chi}=\sum_c \sqrt{p_c}\ket{c}$, one obtains
\begin{equation}
r_{cc'}=(1-\mu)p_c\,\delta_{cc'}+\mu\sqrt{p_cp_{c'}}.
\end{equation}
In particular, $r_{+-}=r_{-+}=\mu\sqrt{p_+p_-}$. A direct calculation using Eq.~\eqref{eq:U_shift} then yields
\begin{align}
\Phi_\mu(\rho_0)&=p_+\,S_+\rho_0 S_+^\dagger + p_0\,\rho_0 + p_-\,S_-\rho_0 S_-^\dagger \nonumber\\
&+ p_-\,P_0\rho_0 P_0 + \mu\sqrt{p_+p_-}\,\big(S_+\rho_0 P_0 + P_0\rho_0 S_-\big).
\label{eq:Phi_full}
\end{align}
The first line reproduces the classical birth--death--lazy dynamics, while the second contains boundary and coherence-induced corrections. In particular, coherence between upward and downward channels, controlled by the parameter $\mu$, gives rise to genuinely quantum contributions. The term $p_- P_0 \rho_0 P_0$ arises from the boundary action of the lowering operator at $n=0$, where transitions to negative energy levels are suppressed. It is convenient to separate the map into incoherent and coherent parts,
\begin{equation}
\Phi_\mu(\rho_0)=\Phi_0(\rho_0)+\Phi_{\mathrm{coh}}(\rho_0),
\end{equation}
where $\Phi_0$ is the classical contribution defined in Eq.~\eqref{eq:Phi_0}, and
\begin{equation}
\Phi_{\mathrm{coh}}(\rho_0)=\mu\sqrt{p_+p_-}\big(S_+\rho_0 P_0 + P_0\rho_0 S_-\big).
\end{equation}
This decomposition shows that, within the present interaction model, only coherence between the channels $\ket{+}$ and $\ket{-}$ contributes to deviations from the classical dynamics.

An immediate consequence of Eq.~\eqref{eq:Phi_full} is that the evolution of populations is independent of the coherence parameter $\mu$. Indeed, for states diagonal in the energy basis, $\rho_0=\sum_{n\ge0}\wp(n)\ket{n}\!\bra{n}$, the terms $S_+\rho_0 P_0$ and $P_0\rho_0 S_-$ are strictly off-diagonal and therefore do not affect the populations. One then obtains
\begin{equation}
\wp'(0)=(p_-+p_0)\wp(0)+p_-\wp(1),
\end{equation}
and, for $n\ge1$,
\begin{equation}
\wp'(n)=p_+\wp(n-1)+p_0\wp(n)+p_-\wp(n+1),
\end{equation}
which coincides with the classical birth--death--lazy process of Sec.~\ref{sec:classical_model}.

Thus, the parameter $\mu$ controls only the generation of quantum coherence, without affecting the marginal energy distribution. In particular, even if the initial state is diagonal, coherence between $\ket{0}$ and $\ket{1}$ is dynamically generated, while the populations evolve classically. This separation between classical population dynamics and coherence generation constitutes the key structural feature of the model.

Importantly, this decoupling is not restricted to diagonal initial states and does not rely on the particular choice of ancillary state $\rho_\mu$. It follows from the structure of the interaction encoded in $U_\mathcal{S}$: the coherence-induced terms in Eq.~\eqref{eq:Phi_full} are strictly off-diagonal in the energy basis and therefore do not contribute to the evolution of populations. As a result, the diagonal elements of $\rho_t$ are fully determined by the classical part of the map, while coherence affects only the off-diagonal sector. In this sense, the separation between population equilibration and coherence dynamics is a structural property of the present evolution. Similar decompositions have been shown to carry clear physical meaning in open quantum systems, where population and coherence sectors give rise to distinct dynamical and thermodynamic behavior \cite{Santos2019}. A detailed analysis of general coherent initial states is left for future work.

\subsection{Equilibration and role of coherence}
\label{subsec:equilibration}

The explicit form of the reduced map $\Phi_\mu$ reveals a structural separation between population dynamics and coherence generation. In particular, when expressed in the energy eigenbasis, $\Phi_\mu$ acts blockwise on diagonal and off-diagonal sectors, with no back-action from coherence onto populations. As a result, the evolution of energy populations depends exclusively on the classical component $\Phi_0$, independently of the coherence parameter $\mu$.

Concretely, denoting $\wp_k(n)=\bra{n}\Phi_\mu^k(\rho_0)\ket{n}$, one finds
\[
\wp_{k+1}(0)=(p_-+p_0)\wp_k(0)+p_-\wp_k(1),
\]
and, for $n\ge1$,
\[
\wp_{k+1}(n)=p_+\wp_k(n-1)+p_0\wp_k(n)+p_-\wp_k(n+1),
\]
which coincides exactly with the classical birth--death--lazy process introduced in Sec.~\ref{sec:classical_model}. Consequently, all statements regarding equilibration, existence of stationary states, and convergence of populations follow directly from classical Markov-chain theory \cite{Norris1997,Levin2009} and are completely insensitive to the presence of coherence.

In this sense, equilibration in energy space is purely classical: whenever the associated stochastic dynamics is ergodic and admits a unique stationary distribution, the populations converge to it from arbitrary initial conditions. Under suitable balance conditions, this stationary distribution takes a Gibbs form. Importantly, this equilibration occurs irrespective of the coherence parameter $\mu$, since the diagonal sector of the dynamics is closed under $\Phi_\mu$.

By contrast, the evolution of off-diagonal elements exhibits genuinely quantum behavior. As seen in Eq.~\eqref{eq:Phi_full}, coherence is generated through terms proportional to $\mu\sqrt{p_+p_-}$, which couple the ground state to the first excited level via the boundary projector $P_0$. Even for initial states diagonal in the energy basis, coherence between $\ket{0}$ and $\ket{1}$ is produced after a single step and is subsequently transported along the ladder by the classical shift dynamics. More generally, any localized breaking of translational invariance in energy space would play an equivalent structural role, independently of the specific realization of the boundary at $n=0$.

This structure exemplifies a broader class of effective dynamics in energy space: channels for which the diagonal restriction defines an ergodic classical transport process, while coherence is injected locally and propagated without influencing populations. Within such dynamics, equilibration of energy distributions does not constrain the long-time behavior of the full density operator. In particular, convergence of populations to a Gibbs distribution does not imply convergence of the quantum state itself.

The present collision-assisted quantum walk provides a minimal realization of this mechanism. While the diagonal sector relaxes irreversibly toward its classical stationary state, the off-diagonal sector retains a nontrivial dependence on $\mu$, leading to persistent coherence. This separation between classical equilibration and quantum state convergence underlies the phenomenon of thermalization without state convergence explored in the following sections.

\subsection{Perturbative regime and persistence of nonequilibrium coherence}
\label{subsec:perturbative_coherence}

In this subsection we specialize to parameter regimes for which the stationary state of the classical dynamics is Gibbsian, so that the stationary diagonal state satisfies $\rho_\infty=\rho_\beta$, as discussed in Sec.~\ref{sec:temporal_equilibrium}. This allows us to directly assess convergence properties with respect to the Gibbs state. In this regime, the instantaneous Gibbs state $\rho_{\beta_t}$ converges to the stationary Gibbs state $\rho_\beta$ as $t\to\infty$, so that both descriptions coincide in the long-time limit.

Further insight into the role of coherence can be obtained by treating the parameter $\mu$ as perturbatively small. From Eq.~\eqref{eq:Phi_full}, the reduced dynamics can be written as
\begin{equation}
\Phi_\mu=\Phi_0+\mu\,\Xi ,
\end{equation}
where $\Phi_0$ denotes the classical birth--death--lazy map and $\Xi$ encodes the coherence-generating contribution localized at the lowest energy levels. Iteration of the map yields, to first order in $\mu$,
\begin{equation}
\Phi_\mu^t=\Phi_0^t+\mu\sum_{j=0}^{t-1}\Phi_0^{\,t-1-j}\,\Xi\,\Phi_0^{\,j}+O(\mu^2),
\end{equation}
which clearly separates classical relaxation in energy space from the leading quantum correction.

For an initial state diagonal in the energy basis,
\begin{equation}
\rho_0=\sum_{n\ge0}\wp_0(n)\ket{n}\!\bra{n},
\end{equation}
the incoherent dynamics satisfies $\Phi_0^t(\rho_0)\to\rho_\beta$. In the presence of coherence ($\mu>0$), however, the first-order correction continuously generates off-diagonal terms. In particular, even when the initial state is chosen to be thermal, one finds
\begin{equation}
\Phi_\mu(\rho_\beta)=\rho_\beta+\mu\sqrt{p_+p_-}\,\wp_\beta(0)\bigl(\ket{1}\!\bra{0}+\ket{0}\!\bra{1}\bigr)+O(\mu^2),
\end{equation}
showing that the Gibbs state is not a fixed point of $\Phi_\mu$ for any $\mu>0$. Since the Gibbs state is diagonal in the energy eigenbasis, this asymptotic deviation is entirely supported by off-diagonal contributions generated by the coherent part of the dynamics.

As a consequence, while the populations converge to their Gibbs distribution independently of $\mu$, the full quantum state does not converge to $\rho_\beta$ unless $\mu=0$. Instead, the dynamics approaches a stationary quantum state whose diagonal sector coincides with $\rho_\beta$ but which retains persistent coherence, leading to a finite asymptotic deviation from the Gibbs manifold.

This behavior can be quantified using the perturbative bounds derived in Appendix~\ref{app:upper-bound}. Denoting by
\begin{equation}
d_{\mathrm{cl}}(t)\equiv d_1\!\bigl(\Phi_0^t(\rho_0),\rho_{\beta_t}\bigr)
\end{equation}
the thermal distance obtained by evolving only under the incoherent map $\Phi_0$, one finds
\begin{equation}
d_{\mathrm{cl}}(t)\le d_{\mathrm{th}}(t)\le
d_{\mathrm{cl}}(t)+\mu\sqrt{p_+p_-}\sum_{j=0}^{t-1}\wp_j(0)+O(\mu^2),
\end{equation}
where $d_{\mathrm{th}}(t)$ is the thermal distance defined in Eq.~\eqref{eq:dth(t)}. Since $d_{\mathrm{cl}}(t)\to0$ as $t\to\infty$, the long-time behavior of $d_{\mathrm{th}}(t)$ is entirely determined by the coherence-induced contribution.

Using the exponential convergence of the classical birth--death--lazy process, the cumulative boundary occupation satisfies
\begin{equation}
\sum_{j=0}^{t-1}\wp_j(0)=t\,\wp_\beta(0)+O(1),
\end{equation}
as shown in Appendix~\ref{app:boundary_sum}. Although the perturbative upper bound grows linearly in time, it is not asymptotically tight. Instead, the full dynamics converges to a stationary quantum state at a finite distance from the Gibbs state. In particular, for sufficiently small $\mu$,
\begin{equation}
d_\infty(\mu)\equiv\lim_{t\to\infty} d_{\mathrm{th}}(t)=O(\mu),
\end{equation}
demonstrating that the breakdown of state convergence is perturbatively controlled.

Taken together, these results establish a minimal and quantitatively bounded scenario of thermalization without state convergence. Classical transport in energy space ensures equilibration of populations, while coherence generated locally at low energies produces a persistent nonequilibrium correction to the quantum state. The magnitude of this deviation is governed jointly by the coherence parameter $\mu$ and by the transport properties of the underlying classical dynamics.

\section{Numerical results}
\label{sec:numerics}

In this section we illustrate the analytical results obtained above through numerical simulations of the reduced dynamics generated by the map $\Phi_\mu$. Since the population dynamics is independent of $\mu$ and coincides with the classical birth--death--lazy process, our numerical analysis focuses on the genuinely quantum contribution of the model, namely, the coherence-induced deviation from thermal relaxation at the level of the full density operator.

The simulations are performed on a truncated ladder $n=0,1,\dots,20$, with transition probabilities $p_+=0.2$, $p_0=0.1$, and $p_-=0.7$. The initial state is taken to be diagonal in the energy basis, given by a normalized discrete Gaussian $\wp_0(n)\propto e^{-(n-n_0)^2/(2\sigma^2)}$ with $n_0=2$ and $\sigma=2$.

\begin{figure}
\centering
\includegraphics[scale=0.5]{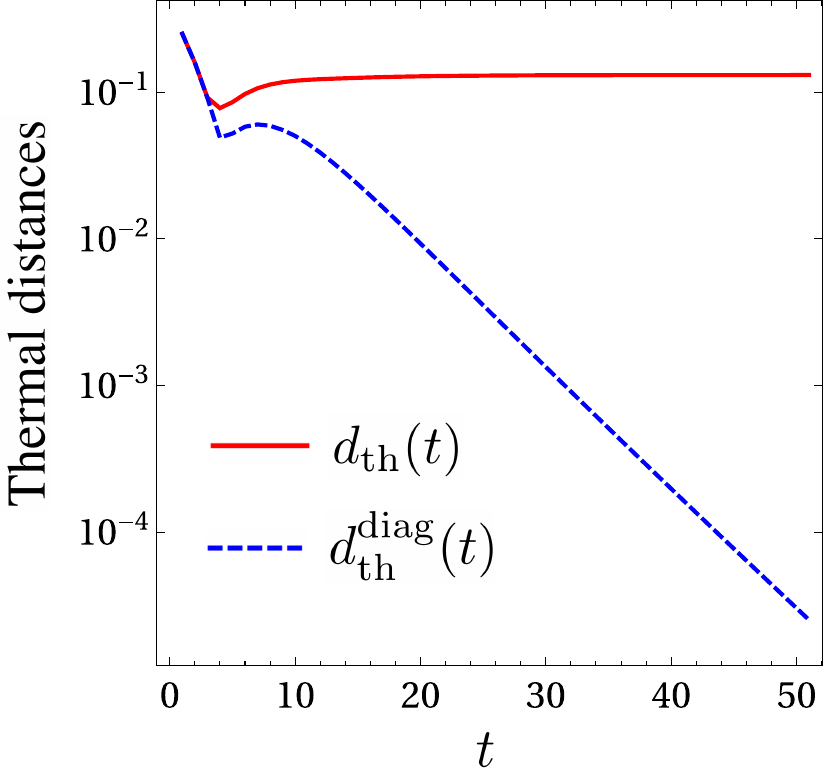}
\caption{Thermal distance $d_{\mathrm{th}}(t)$ (solid red line) and its dephased counterpart $d_{\mathrm{th}}^{\mathrm{diag}}(t)$ (dashed blue line), where $\rho_{\beta_t}$ is the Gibbs state with the same instantaneous mean energy as $\rho_t$. While the dephased dynamics relaxes to the Gibbs state, $d_{\mathrm{th}}(t)$ saturates at a finite value, revealing a persistent coherence-induced deviation from thermal equilibrium. Parameters as described in the text.}
\label{fig2}
\end{figure}

To isolate the origin of the deviation from thermal behavior, we first compare the full thermal distance 
$d_{\mathrm{th}}(t)=d_1(\rho_t,\rho_{\beta_t})$ 
with its dephased counterpart 
$d_{\mathrm{th}}^{\mathrm{diag}}(t)=d_1(\mathrm{diag}(\rho_t),\rho_{\beta_t})$. 
As shown in Fig.~\ref{fig2}, the difference between the two curves directly quantifies the contribution of coherence, confirming that the departure from thermal behavior arises exclusively from off-diagonal terms. While the dephased dynamics reproduces classical relaxation towards the Gibbs state, the full distance remains finite due to the coherence generated at the boundary.

\begin{figure}
\centering
\includegraphics[scale=0.5]{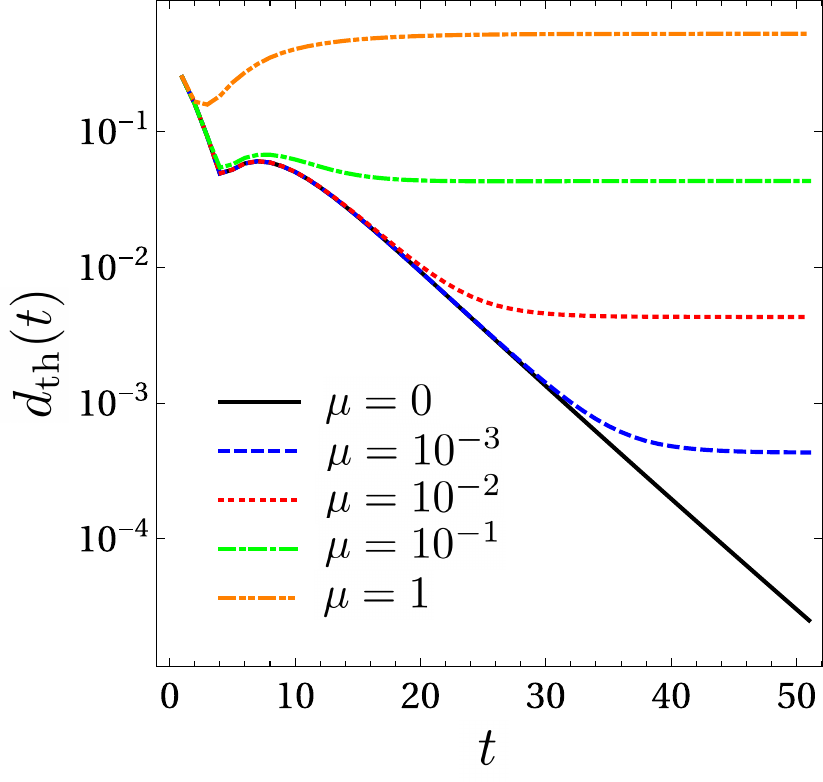}
\caption{Time evolution of the thermal distance $d_{\mathrm{th}}(t)$ for different values of the coherence parameter $\mu$. For $\mu=0$, the dynamics reduces to the classical case and $d_{\mathrm{th}}(t)\to0$, while for $\mu>0$ the distance saturates at a finite value that increases with $\mu$, indicating a growing coherence-induced deviation from the Gibbs state. Parameters as described in the text.}
\label{fig3}
\end{figure}

We then analyze how this effect depends on the coherence parameter $\mu$. As shown in Fig.~\ref{fig3}, the short-time dynamics remains close to the classical behavior for all values of $\mu$, reflecting the fact that coherence is initially negligible. At longer times, however, a clear separation emerges, with increasing $\mu$ leading to larger asymptotic deviations from the thermal state. This behavior highlights that coherence does not affect equilibration at the level of populations, but modifies the structure of the quantum state in a persistent way.

\begin{figure}
\centering
\includegraphics[scale=0.5]{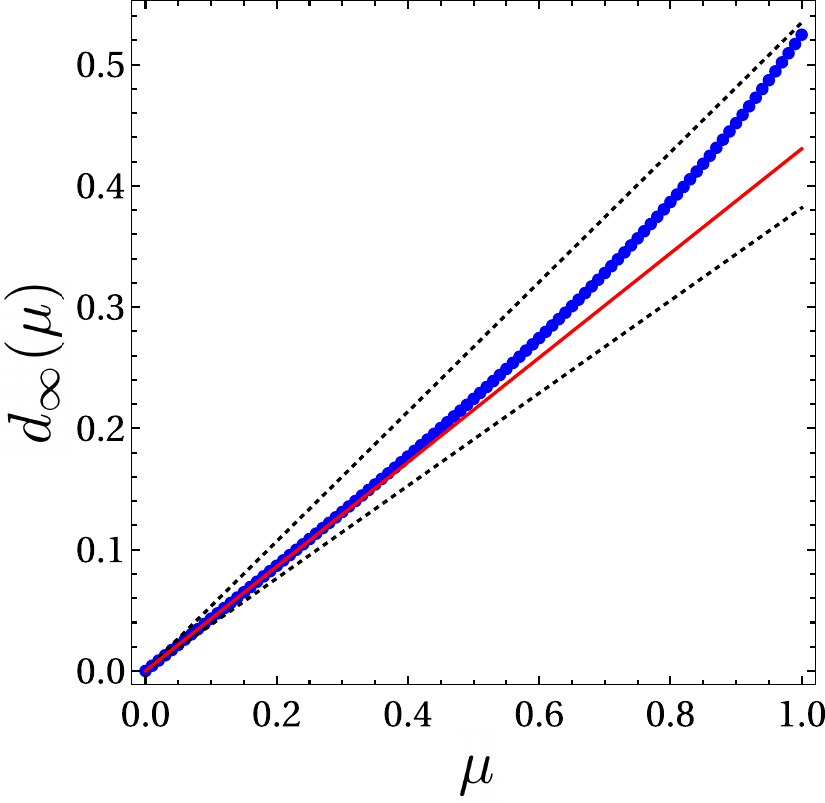}
\caption{Asymptotic thermal distance $d_\infty(\mu)$ as a function of the coherence parameter $\mu$. Blue dots denote numerical results, dashed black lines indicate analytical bounds, and the red solid line corresponds to the first-order perturbative prediction. The linear scaling at small $\mu$ confirms the perturbative analysis, while deviations at larger $\mu$ signal the onset of nonperturbative effects. Parameters as described in the text.}
\label{fig4}
\end{figure}

Finally, we consider the asymptotic quantity 
$d_\infty(\mu)=\lim_{t\to\infty} d_{\mathrm{th}}(t)$, 
which provides a direct quantitative measure of the coherence-induced deviation from the Gibbs manifold. As shown in Fig.~\ref{fig4}, $d_\infty(\mu)$ grows linearly for small $\mu$, in agreement with the perturbative analysis, while clear deviations from linearity appear at larger values of $\mu$. Importantly, the numerical results remain fully confined within the analytical bounds, confirming that the effect of coherence is both nontrivial and quantitatively controlled.

\section{Conclusions}
\label{sec:conclusions}

In this work we introduced energy-space quantum walks as a minimal and flexible framework to investigate equilibration, thermalization, and irreversibility from the perspective of effective dynamics. By reinterpreting the configuration space of a walk as a ladder of energy eigenlevels, thermalization is naturally framed as transport in energy space, allowing its key features to be analyzed independently of microscopic system--bath models.

At the classical level, we showed that a simple birth--death--lazy dynamics on an energy ladder already captures essential aspects of equilibration. In particular, nonunitality and bias in energy space give rise to irreversible transport and to the emergence of unique stationary states. Under suitable balance conditions, these stationary distributions acquire a Gibbs form, demonstrating that Gibbsian statistics can arise as a property of effective energy transport rather than from an explicit thermal reservoir.

We then embedded this stochastic dynamics into a unitary, collision-assisted quantum walk by coupling the system to ancillary degrees of freedom that encode elementary energy-exchange channels. This construction provides a transparent interpolation between incoherent and coherent regimes, controlled by a single parameter. A central structural result of this embedding is the complete decoupling between population dynamics and coherence generation: while energy populations evolve exactly as in the associated classical Markov process, quantum coherence is injected locally and propagated without back-action on the diagonal sector.

This separation leads to a minimal and quantitatively controlled scenario of thermalization without state convergence. Whenever the diagonal sector induces ergodic transport in energy space, the populations equilibrate and, when detailed balance holds, relax to a Gibbs distribution. However, for any finite amount of coherence, the full quantum state does not converge to the Gibbs state. Instead, it approaches a stationary quantum state whose diagonal coincides with the thermal distribution but which retains persistent coherence. A perturbative analysis shows that the resulting deviation from the Gibbs manifold is finite in the long-time limit and scales linearly with the coherence parameter in the weak-coherence regime, with its magnitude governed by the classical transport properties near the low-energy boundary.

More broadly, our results highlight that thermalization is not a monolithic notion: equilibration of energy distributions and convergence of quantum states can be structurally distinct phenomena. From this viewpoint, energy-space quantum walks provide a useful language to disentangle classical transport mechanisms from genuinely quantum corrections and to identify minimal conditions under which coherence survives equilibration.

Although our analysis was formulated in terms of a collision-assisted quantum walk, the underlying mechanism is not intrinsically tied to this specific realization. Closely related phenomena may naturally arise in more traditional discrete-time quantum walks with an explicit coin degree of freedom, where coherence generated by the coin can similarly decouple from population transport along the energy ladder. In this sense, standard coin-based quantum walks offer a promising and experimentally relevant arena to explore coherence-controlled transport and nonequilibrium effects in energy space.

Several directions for future research follow from this framework. One promising avenue is the formulation of energy-conserving embeddings in which the ancillary or coin degrees of freedom carry an explicit energetic interpretation, allowing for a consistent thermodynamic accounting of work and heat. Another direction concerns the exploration of continuous-time limits and their relation to Lindblad-type generators, clarifying when energy-space dynamics can or cannot be reduced to standard open-system descriptions. Extensions to more complex energy graphs or interacting ladders may further enrich the interplay between coherence, transport, and irreversibility. Taken together, these perspectives suggest that transport in energy space offers a powerful and conceptually transparent route to understanding thermal behavior and its genuinely quantum deviations.

\section*{Acknowledgments}
This study was financed in part by the Coordenação de Aperfeiçoamento de Pessoal de Nível Superior -- Brasil (CAPES) -- Finance
Code 001. R.~M.~A.\ acknowledges support from the Conselho Nacional de Desenvolvimento Científico e Tecnológico (CNPq), Grant No.~305957/2023-6.

\section*{Data availability}
The code used to generate the results and figures of this study is publicly available~\cite{angelo2026code}.

\appendix

\section{Derivation of the stationary distribution for the birth--death--lazy dynamics}
\label{app:stationary_derivation}

In this appendix we derive the stationary distribution of the birth--death--lazy dynamics defined in Sec.~\ref{sec:classical_model}. As shown there, stationarity implies the local balance condition
\begin{equation}
p_-\,\wp_\infty(n+1)=p_+\,\wp_\infty(n),\qquad n\ge0 .
\label{eq:appendix_balance}
\end{equation}
We now derive the same result directly from the recurrence relation.

At stationarity the populations satisfy
\begin{equation}
\wp_\infty(n)=p_-\,\wp_\infty(n+1)+p_0\,\wp_\infty(n)+p_+\,\wp_\infty(n-1),
\label{eq:appendix_recurrence_full}
\end{equation}
with $n\ge 1$. Rearranging and using $p_++p_-+p_0=1$ gives
\begin{equation}
(p_++p_-)\wp_\infty(n)=p_-\,\wp_\infty(n+1)+p_+\,\wp_\infty(n-1).
\label{eq:appendix_recurrence}
\end{equation}
We solve Eq.~\eqref{eq:appendix_recurrence} by means of the exponential ansatz $\wp_\infty(n)\propto \lambda^n$. Substitution yields the characteristic equation $p_-\,\lambda^2-(p_++p_-)\lambda+p_+=0 $, whose roots are
\begin{equation}
\lambda_\pm=\frac{(p_++p_-)\pm\sqrt{(p_++p_-)^2-4p_-p_+}}{2p_-}.
\end{equation}
Since $(p_++p_-)^2-4p_-p_+=(p_--p_+)^2$, we obtain
\begin{equation}
\lambda_\pm=\frac{(p_++p_-)\pm|p_--p_+|}{2p_-}.
\end{equation}
For the biased case $p_->p_+$ relevant to equilibration, these roots simplify to
\begin{equation}
\lambda_+=1,\qquad\lambda_-=\frac{p_+}{p_-}.
\end{equation}
The general solution of the homogeneous recurrence \eqref{eq:appendix_recurrence} thus takes the form
\begin{equation}
\wp_\infty(n)=A+B\left(\frac{p_+}{p_-}\right)^n,
\label{eq:general_solution}
\end{equation}
with constants $A,B\in\mathbb{R}$. Normalizability of $\wp_\infty(n)$ on $n\in\mathbb{N}$ requires $\sum_{n\ge 0}\wp_\infty(n)<\infty$.
Since the constant term $A$ leads to a divergent series, one must set $A=0$. This leaves the geometric form
\begin{equation}
\wp_\infty(n)=B\left(\frac{p_+}{p_-}\right)^n,
\label{eq:appendix_geometric}
\end{equation}
which is normalizable if and only if $p_+/p_-<1$, i.e., $p_->p_+$. The normalization constant $B=\wp_\infty(0)$ is fixed by imposing $\sum_{n\ge 0}\wp_\infty(n)=1$, yielding
\begin{equation}
\wp_\infty(0)=\left(\sum_{n=0}^{\infty} \left(\frac{p_+}{p_-}\right)^n
\right)^{-1}=\frac{p_- - p_+}{p_-}.
\end{equation}
This completes the derivation of the stationary distribution reported in Sec.~\ref{sec:classical_model}.

It is worth noting that the stationary distribution depends only on the ratio $p_+/p_-$ and is therefore independent of the lazy probability $p_0$. The stay channel affects the kinetics of the dynamics---in particular the relaxation times---but cancels out of the equilibrium balance condition. This separation between kinetic rates and stationary weights reflects a general property of birth--death--lazy processes with self-loops.

\section{Upper bound on the thermal distance}
\label{app:upper-bound}

We derive an upper bound for the trace distance
$d_{\mathrm{th}}(t) =\frac{1}{2} \|\rho_t - \rho_\beta\|_1$,
using the perturbative expansion
\begin{equation}
\rho_t=\Phi_0^t(\rho_0)+\mu\,\Sigma_t(\rho_0)+O(\mu^2).
\end{equation}
By the triangle inequality,
\begin{equation}
d_{\mathrm{th}}(t)\le\frac{1}{2}\|\Phi_0^t(\rho_0)-\rho_\beta\|_1+
\frac{\mu}{2}\,\|\Sigma_t(\rho_0)\|_1+O(\mu^2).
\end{equation}
The first term corresponds to the classical contribution,
\begin{equation}
\frac{1}{2}\|\Phi_0^t(\rho_0)-\rho_\beta\|_1 = d_{\mathrm{cl}}(t).
\end{equation}
To bound the second term, we use the explicit form
\begin{subequations}
\begin{align}
\Sigma_t &=\sqrt{p_+p_-}\sum_{j=0}^{t-1}
\wp_j(0)\,\Phi_0^{\,t-1-j}(A),\\
A &= \ket{1}\!\bra{0} + \ket{0}\!\bra{1}.
\end{align}
\end{subequations}
which follows from the repeated evolution of coherences under the dynamics.
Applying the triangle inequality,
\begin{equation}
\|\Sigma_t\|_1
\le
\sqrt{p_+p_-}
\sum_{j=0}^{t-1}
\wp_j(0)\,
\|\Phi_0^{\,t-1-j}(A)\|_1.
\end{equation}
Since $\Phi_0$ is a CPTP map, it is contractive with respect to the trace norm on Hermitian operators. Therefore, for any Hermitian operator $X$,
$\|\Phi_0^k(X)\|_1 \le \|X\|_1$, so that
\begin{equation}
\|\Sigma_t\|_1\le \sqrt{p_+p_-} \sum_{j=0}^{t-1} \wp_j(0)\,\|A\|_1.
\end{equation}
A direct calculation yields $\|A\|_1 = 2$, hence
\begin{equation}
\|\Sigma_t\|_1\le 2\sqrt{p_+p_-} \sum_{j=0}^{t-1}\wp_j(0).
\end{equation}
Combining the above results, we obtain
\begin{equation}
d_{\mathrm{th}}(t)
\le
d_{\mathrm{cl}}(t)
+
\mu\,\sqrt{p_+p_-}
\sum_{j=0}^{t-1}\wp_j(0)
+
O(\mu^2).
\end{equation}

\section{Asymptotic behavior of the boundary occupation}
\label{app:boundary_sum}

In this appendix we derive the asymptotic behavior of the cumulative boundary occupation,
\begin{equation}
\sum_{j=0}^{t-1}\wp_j(0),
\end{equation}
which enters the perturbative bounds discussed in Sec.~\ref{subsec:perturbative_coherence}. For the classical birth--death--lazy dynamics with $p_->p_+$, the Markov chain is ergodic and converges exponentially fast to the stationary distribution $\wp_\infty(n)$ (see, e.g., Refs.~\cite{Norris1997,Levin2009}). In particular, there exist constants $C>0$ and $\Delta>0$ such that
\begin{equation}
\big|\wp_t(0)-\wp_\infty(0)\big|\le C e^{-\Delta t}.
\label{eq:boundary_exp_conv}
\end{equation}

The cumulative sum can be written as  
\begin{align}
\sum_{j=0}^{t-1}\wp_j(0)
&=\sum_{j=0}^{t-1}\wp_\infty(0)
+\sum_{j=0}^{t-1}\big(\wp_j(0)-\wp_\infty(0)\big)
\nonumber\\
&=t\,\wp_\infty(0)+
\sum_{j=0}^{t-1}\big(\wp_j(0)-\wp_\infty(0)\big).
\end{align}
The second term is bounded uniformly in $t$, since
\begin{equation}
\left|\sum_{j=0}^{t-1}\big(\wp_j(0)-\wp_\infty(0)\big)\right|
\le\sum_{j=0}^{\infty} C e^{-\Delta j}
=\frac{C}{1-e^{-\Delta}}.
\end{equation}
Therefore,
\begin{equation}
\sum_{j=0}^{t-1}\wp_j(0)
=t\,\wp_\infty(0)+O(1),
\label{eq:boundary_sum_result}
\end{equation}
which establishes the claimed asymptotic behavior.

\bibliography{references}

\end{document}